\documentclass{elsart}
\usepackage[dvips]{graphics,epsfig,color}
\usepackage{amssymb}
\newcommand{\beq}{\begin{equation}}
\newcommand{\eeq}{\end{equation}}
\newcommand{\bqq}{\begin{eqnarray}}
\newcommand{\eqq}{\end{eqnarray}}

\usepackage{graphicx}
\usepackage{dcolumn}
\usepackage{bm}
\usepackage{amsmath}
\usepackage{psfrag}
\usepackage{color}
\usepackage{latexsym,amsfonts}
\usepackage{graphicx}
\providecommand{\href}[2]{#2}   
\definecolor{Blue2}{rgb}{0.,0.,0.8125}
\definecolor{Brown3}{rgb}{0.625,0.25,0.}
\definecolor{Cyan4}{rgb}{0.,0.56,0.56}
\definecolor{Green4}{rgb}{0.,0.56,0.}
\definecolor{LtBlue}{rgb}{0.27,0.42,0.52}
\definecolor{Magenta4}{rgb}{0.5625,0.,0.5625}
\definecolor{Red2}{rgb}{0.8125,0.,0.}
\newcommand{\XXi}{\Xi_{3/2}}

\begin{document}
\runauthor{D.L.~Borisyuk}
\begin{frontmatter}
\title{Spin and Parity of $\XXi$ Exotic Baryon from Kaon Scattering on the Nucleon
}
\author{D.L.~Borisyuk},
\author{A.P.~Kobushkin}\footnote{email address: kobushkin@bitp.kiev.ua} and
\author{Yu.V.~Kutafin}\footnote{email address: kutafin@i.com.ua}
\address{Bogolyubov Institute for Theoretical Physics,
Metrologicheskaya str. 14B, 03143 Kiev, Ukraine
}
\begin{abstract}
We calculate total cross section for production of the  $\XXi(1862)$ exotic baryon in $\bar{K}N \to K\XXi$ reaction  assuming the following spin-parity values of the $\XXi$ baryon $J^\pi=\frac12^+$, $\frac12^-$, $\frac32^+$  and $\frac32^-$.  We demonstrate that the reaction total cross section strongly depends on the spin of the $\XXi$ baryon.
\end{abstract}
\noindent
\begin{keyword}
exotic baryon, spin, parity
\end{keyword}
\end{frontmatter}

\date{\today}

\maketitle

\section{Introduction\label{sec.1}}
Experimental observation \cite{Nakano,Barmin,Stepanyan,Barth,Asratyan,Kubarovsky,Togoo,Aleev,Airapetian,Abdel-Bary,Aslanyan,Troyan,ZEUS} of a narrow baryon, $\Theta^+$, which, due to its strangness $S=+1$, cannot be a three-quark bound system pushes great interest to physics of exotic hadrons, see, {\it e.g.}, \cite{Zhu,Oka} and further references therein. The $\Theta^+$ mass is close to $1540~\mathrm{MeV/c^2}$ and width is much smaller than a typical hadron width. It was found no evidence for $\Theta^{++}$ \cite{Barth,Kubarovsky,Airapetian,ZEUS} which leads to the conclusion that $\Theta^+$ should be isoscalar. The $\Theta^+$ baryon has been included as a three-star resonance in 2004 PDG listings.

Somewhat later two another candidates for the exotic baryons, $\XXi^{--}$ and $\XXi^{0}$, with strangeness $S=-2$, mass near $1860~\mathrm{MeV/c^2}$ and narrow width  $< 18~\mathrm{MeV}$ were reported by the NA49 collaboration \cite{Alt}. According to its strangness and electric charge the minimal value of the $\XXi$ isospin is $I=\frac32$. It is natural to assume that the isospin singlet $\Theta^+$ and isospin quartet $\XXi^{--}$,  $\XXi^{-}$, $\XXi^{0}$ and  $\XXi^{+}$ should be members of the  same flavor  multiplet ($\mathbf{ \bar{10}}_f$).
   
Besides that, a narrow anti-charmed baryon with a minimal constituent quark composition $uudd\bar{c}$ was observed by the H1 collaboration \cite{Aktas}.

Finally, a narrow peak at 1734~$\mathrm{MeV/c^2}$ in $\Lambda K^0_s$ invariant mass observed in preliminary results from the STAR experiment at RHIC was interpreted as a pentaquark state with the isospin $I=\frac12$ \cite{Kabana}.

Despite these impressive results, the negative results of a search for $\Theta^+$ \cite{Bai,Knopfle,Pinkenburg,Antipov}, as well as for $\XXi$ \cite{Adamovich,Fischer}, were also reported very recently. So it is evident that a new kind of experimental study is necessary to clarify the situation. For example,  
\begin{itemize}
\item Experiments with high statistics and different beams and targets to confirm or reject the observed exotic baryons and to search, if any, for new exotic states.
\item Measurement of spin and parity of the observed pentaquarks.  
\end{itemize}

A lot of theoretical models were proposed to interpret the obtained experimental results for the exotic baryons, chiral-skyrmion models, constituent quark models, QCD sum rules, lattice QCD, etc., see discussion in \cite{Zhu,Oka}. Here it is important to stress, that all such calculations, been ``turned to'' the experimental mass and width of the $\Theta^+$, give very different predictions for the spectroscopy of exited exotic baryons, as well as predict different spin-parity quantum numbers for the  $\Theta^+$ and the $\XXi$. For example, in the chiral-skyrmion model the lowest exotic states are members of the $\mathbf{ \bar{10}}_f$-plet with $J^\pi=\frac12^+$ \cite{Praszalowicz,Praszalowicz03,DPP}. The next states belong to the $\mathbf{27}_f$-plet with $J^\pi=\frac32^+$ \cite{WKop,BFK,BinWu,EKP}. Some of them are very close to appropriate states from the $\mathbf{ \bar{10}}_f$-plet, but could have different flavor quantum numbers. In turn, the constituent quark model predicts two partners of ideally mixed $\mathbf{ \bar{10}}_f$ and $\mathbf{8}_f$ multiplets with $J^\pi=\frac12^+$ and $\frac32^+$ splitting within tens of MeV \cite{DudekClose}.

The aim of this paper is to estimate the production cross section of the $\XXi$ baryon with spin $\frac32$. We are concentrated on the simplest strong interaction reaction,  $\XXi$ production in $\bar{K}N$ scattering
\begin{equation}
\label{reaction}
 \bar{K}N\to K\XXi,
\end{equation}
and demonstrate that the total cross section for the exotic baryon with spin $\frac32$ is at least 50 times larger than that for the exotic baryon with spin $\frac12$. 

The paper is organized as follows. In Section~\ref{Effective} we formulate the model. Then in Section~\ref{Numerical} we determine the parameters of the model and provide numerical calculations. Conclusions are given in Section~\ref{Conclusions}.

\begin{figure}
  \centering  
  \includegraphics[height=0.20\textheight]{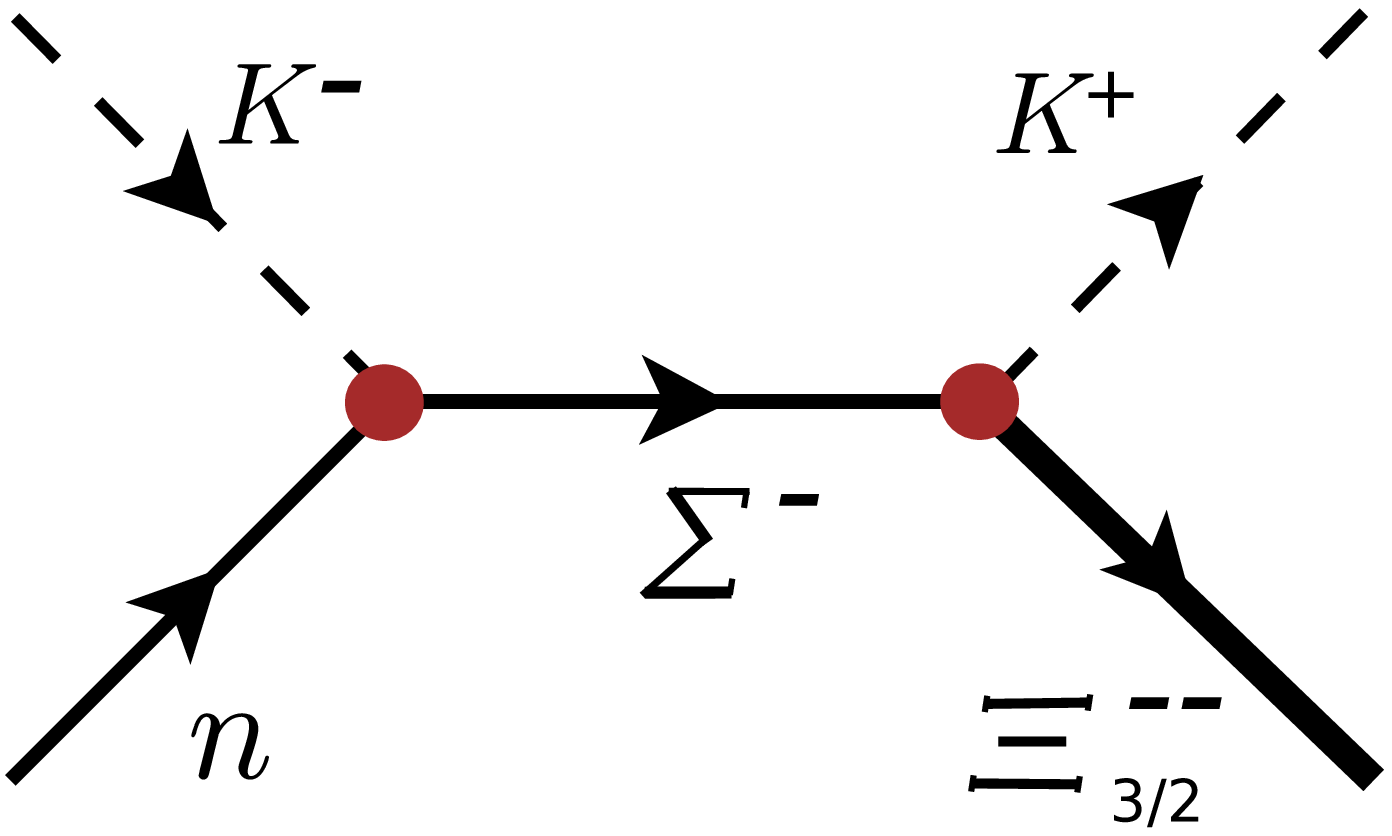}  
  \includegraphics[height=0.24\textheight]{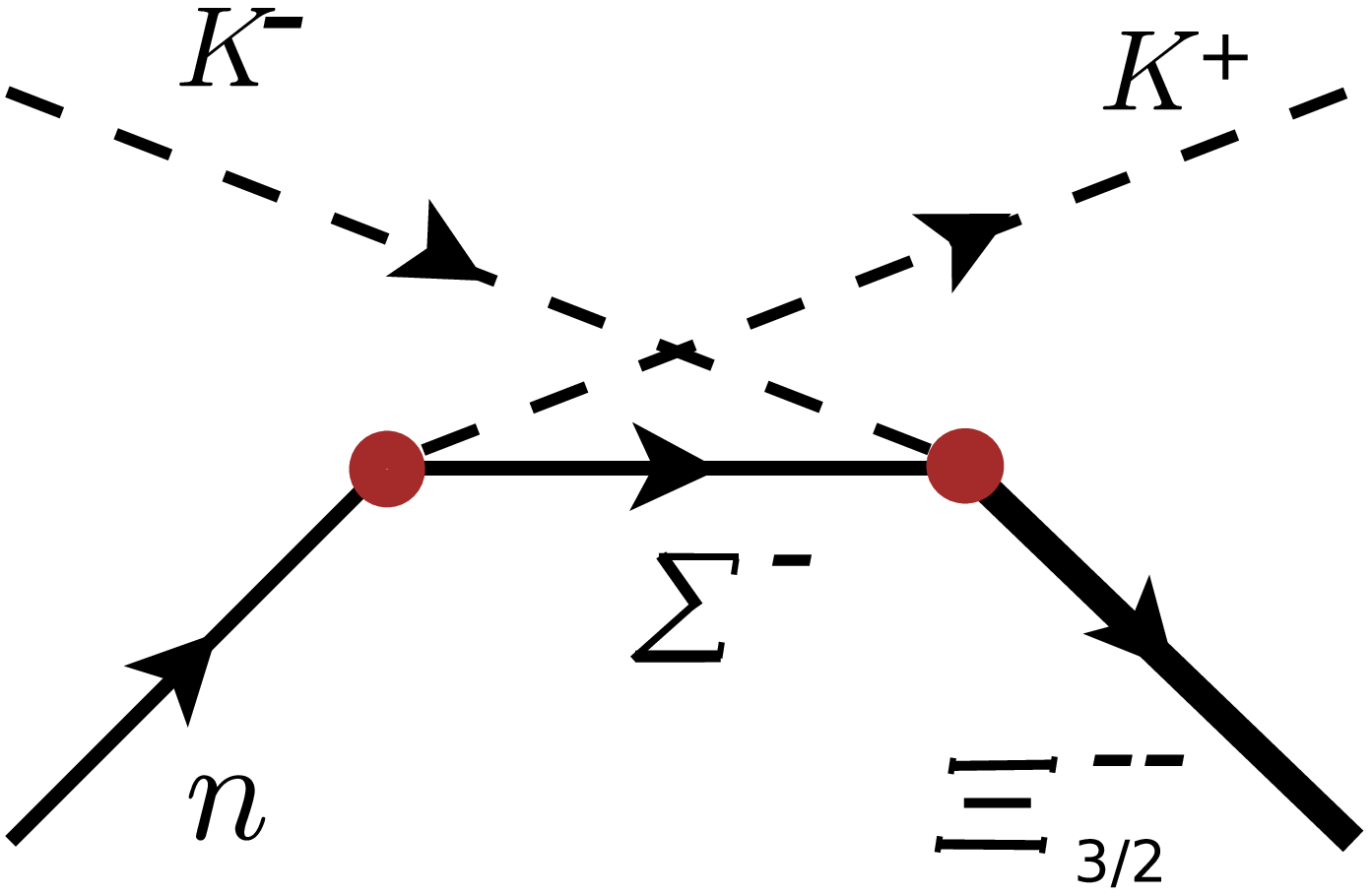}
  \caption{Born diagrams for $\XXi^{--}$ production in $K^-n \to K^+\XXi^{--}$ reaction.}
 \label{fig:diagrams}
\end{figure} 
\section{Model and effective Lagrangians\label{Effective}}
We estimate the reaction cross section considering Born diagrams with $\Sigma$~baryon pole in the $s$ and $u$~channels, Figure~\ref{fig:diagrams}. We use $k$ and $k'$ for the kaon momentum in the initial and final state, $p$ and $q$ for the proton and $\XXi$ momentum, respectively. 

The cross sections for different channels of the reaction (\ref{reaction}) are connected by isospin Clebsh-Gordan coefficients
\begin{equation}
\label{Clebsh-Gordan}
\begin{split}
&\sigma(\bar K^0 p \to K^0\XXi^+)=3\sigma(\bar K^0 p \to K^+\XXi^0)=3\sigma(K^- p \to K^0\XXi^0)=\\
=3&\sigma(\bar K^0 n \to K^+\XXi^-)=3\sigma(\bar K^0 n \to K^0\XXi^0)=3\sigma( K^- p \to K^+\XXi^-)=\\
=3&\sigma(K^- n \to K^0\XXi^-)=\sigma(K^- n \to K^+\XXi^{--}).
\end{split} 
\end{equation}
%

The $KN\Sigma$ effective Lagrangian is well known
\begin{equation}
\label{KNSigma}
\mathcal{L}_{KN\Sigma}=ig_{KN\Sigma}\overline{\Sigma}\gamma_5 K N + \text{h.c.}
\end{equation}
with the coupling constant $g_{KN\Sigma}=3.54$ \cite{NIJM}. From J\"ulich-Bonn potential  $g_{KN\Sigma}=5.38$ \cite{J-Bonn}. We use the first value. Because spin and parity of the $\XXi$~baryon are unknown we use one of the following  $K\Sigma\XXi$ Lagrangians depending on the spin-parity of the $\XXi$~baryon
\begin{eqnarray}
\begin{array}{lll}\label{Xxi_coupling}
\mathcal{L}_{K\Sigma\XXi}=ig_{K\Sigma\XXi}\overline{\Sigma}\gamma_5 K \XXi + \text{h.c.},&\text{for} & J^\pi(\XXi)=\frac12^+ \\
\mathcal{L}_{K\Sigma\XXi}=g_{K\Sigma\XXi}\overline{\Sigma} K \XXi + \text{h.c.},&& J^\pi(\XXi)=\frac12^-\\
\mathcal{L}_{K\Sigma\XXi}=i\frac{g_{K\Sigma\XXi}}{m_{\XXi}}\overline{\Sigma}\gamma_5\partial_\mu K \XXi^\mu + \text{h.c.},&& J^\pi(\XXi)=\frac32^+\\
\mathcal{L}_{K\Sigma\XXi}=\frac{g_{K\Sigma\XXi}}{m_{\XXi}}\overline{\Sigma}\partial_\mu K  \XXi^\mu + \text{h.c.},&& J^\pi(\XXi)=\frac32^-
\end{array}
\end{eqnarray}
 In (\ref{Xxi_coupling}) we use Rarita-Schwinger field  $\XXi^\mu(x)$ for a particle with spin $\frac32$. An additional factor $1/m_{\XXi}$ (where $m_{\XXi}$ is the $\XXi$ mass) is introduced to make coupling constant $g_{K\Sigma\XXi}$  dimensionless for the spin-$\frac32$ $\XXi$~baryon. 

A spinor $U^\mu(q)$ for a free spin-$\frac32$ particle satisfies the following equation
\begin{equation}
\label{Rarita-Schwinger}
( q\!\!\!/ -m_{\XXi})U^\mu(q)=0,
\end{equation}
with constraints
\begin{equation}
\label{constraints}
\gamma_\mu U^\mu(q)=0
\end{equation}
where $ q\!\!\!/ \equiv \gamma_\nu q^\nu$ and $\gamma_\nu$ are Dirac $4 \times 4$ matrices. The normalization condition reads
\begin{equation}
\label{normalization}
\overline{U}^\mu(q) U_\mu(q)=-2 m_{\XXi}.
\end{equation}
The spin summation formula for the Rarita-Schwinger spinor reads
\begin{equation}
\label{spin_sum}
\begin{split}
\sum_\mathrm{spin}U_\mu(q)\overline{U}_\nu(q)=&- {1 \over 3 m_{\XXi}}(q\!\!\!/+m_{\XXi}) \left(
g_{\mu\nu} - {q_\mu q_\nu \over m_\Xi^2} -{1 \over 4}[\gamma_\mu,\gamma_\nu]
\right) (q\!\!\!/+m_{\XXi})\equiv\\
\equiv & P_{\mu\nu}(q).
\end{split}
\end{equation}

Reaction scattering amplitude squared, summed over spin states of the $\XXi$ and averaged over nucleon spin states reads
\begin{equation}\label{Square}
\overline{|\mathcal{M}|^2}=
\overline{|\mathcal{M}_s|^2} +  \overline{\mathcal{M}_s\mathcal{M}_u^\ast} +\overline{\mathcal{M}_s^\ast\mathcal{M}_u} +  \overline{|\mathcal{M}_u|^2},
\end{equation}
where $\mathcal{M}_s$ and $\mathcal{M}_u$ are the $s$ and $u$ pole terms corresponding to left and right diagrams of Figure~\ref{fig:diagrams}. For the $\XXi$~baryon with $J^\pi=\frac32^+$  appropriate terms of (\ref{Square}) are 
\begin{equation}\label{xy-terms}
\begin{split}
\overline{\mathcal{M}_x\mathcal{M}_y^\ast}=&\frac12 g^2_{K\Sigma\XXi}g^2_{KN\Sigma}\frac{F^2(\kappa^2_x)}{\kappa^2_x-m^2_\Sigma}\cdot\frac{F^2(\kappa^2_y)}{\kappa^2_y-m^2_\Sigma}\cdot\frac{\kappa_x^\mu\kappa_y^\nu}{3m^2_{\XXi}} \times \\
& \times \mathrm{Tr}\left\{
\left(\kappa_x \!\!\!\!\!\!/ -m_\Sigma\right)P_{\mu\nu}(q)\left(\kappa_y \!\!\!\!\!\!/ -m_\Sigma\right)\left( p\!\!\!/ + m_N\right)
\right\},
\end{split} 
\end{equation}
where $x$, $y$ labels either $s$- or $u$-channel, $\kappa_s$ ($\kappa_u$) is the momentum of the intermediate $\Sigma$~baryon, $\kappa_s = k+p$, $\kappa_u = q-k$, and $F(\kappa^2)$ is form factor. we use a relativistically invariant parameterization for the form factor from Ref.~\cite{NHK}
\begin{equation}\label{forfactor}
F(\kappa^2)=\frac{\Lambda^2}{\sqrt{\Lambda^4+\left(\kappa^2-m^2_\Sigma\right)^2}}, 
\end{equation}
extracted from the cross section $\gamma p \to K^+ \Lambda$. The cut-off parameter $\Lambda=0.85$~GeV \cite{NHK}. The above formula (\ref{xy-terms}) is for positive parity $\XXi$~baryon, in the case of negative parity one has to change the sign of $\XXi$ mass entering $P_{\mu\nu}$.

The trace in the expression (\ref{xy-terms})  was calculated on computer analytically. 

\section{Numerical calculations and discussion of the results\label{Numerical}}

We estimate the coupling constant $g_{K\Sigma\XXi}$ by the same procedure, which was used in Ref.~\cite{Hosaka}. Both exotic baryons, $\Theta^+$ and $\XXi$ are assumed to belong to the same $\mathrm{SU(3)_f}$ multiplet. Assuming $\mathrm{SU(3)_f}$ symmetry for the interaction one gets
\begin{equation}
\label{SU(3)}
g_{KN\Theta}=g_{K\Sigma\XXi}.
\end{equation}
Since the $\Theta^+$ has only one decay channel, $\Theta^+ \to KN$, one can simply calculate the coupling constant $g_{KN\Theta}$ from the total $\Theta$ width%
\begin{equation}
\label{width}
\Gamma_\Theta=g_{KN\Theta}^2\frac{(m_\Theta \mp m_N)^2 - m_K^2}{4\pi m^2_\Theta}
\times\left\{
\begin{array}{cl}
Q                       & \text{ for } J^\pi=\frac12^\pm\\
\frac{Q^3}{3m^2_\Theta} & \text{ for } J^\pi=\frac32^\pm
\end{array}
\right.
\end{equation}
where $m_\Theta$, $m_N$ and $m_K$ are masses of the $\Theta^+$, the nucleon and the kaon and
\begin{equation}
\label{Q_momentum}
Q=\frac1{2m_\Theta}\sqrt{m_\Theta^4 + m_N^4 +m_K^4 - 2m_\Theta^2m_N^2 - 2m_\Theta^2m_K^2-2m_N^2m_K^2}
\end{equation}
is the kaon momentum in the $\Theta^+$ rest frame. Taking $\Gamma_\Theta=15\ \rm{MeV}$ as an upper limit for the $\Theta^+$ decay width one obtains values summarized in Table~\ref{tab:1}.

\begin{table}[t]
\caption{\label{tab:1}Estimated coupling constant $g_{K\Sigma\XXi}$ for different values of the $\XXi$ spin $J$ and parity $\pi$. The data for $J^\pi=\frac12^\pm$ are from Ref.~\cite{Hosaka}.}
\begin{center}
\begin{tabular}{|c||c|c|}
\hline
     & $J=1/2$ & $J=3/2$ \\
\hline\hline
$\pi=+1$ & 3.84 & 38.4 \\
\hline
$\pi=-1$ & 0.53 & 5.34 \\
\hline
\end{tabular}
\end{center}
\end{table}
\begin{figure}
  \centering  
  \includegraphics[height=0.4\textheight]{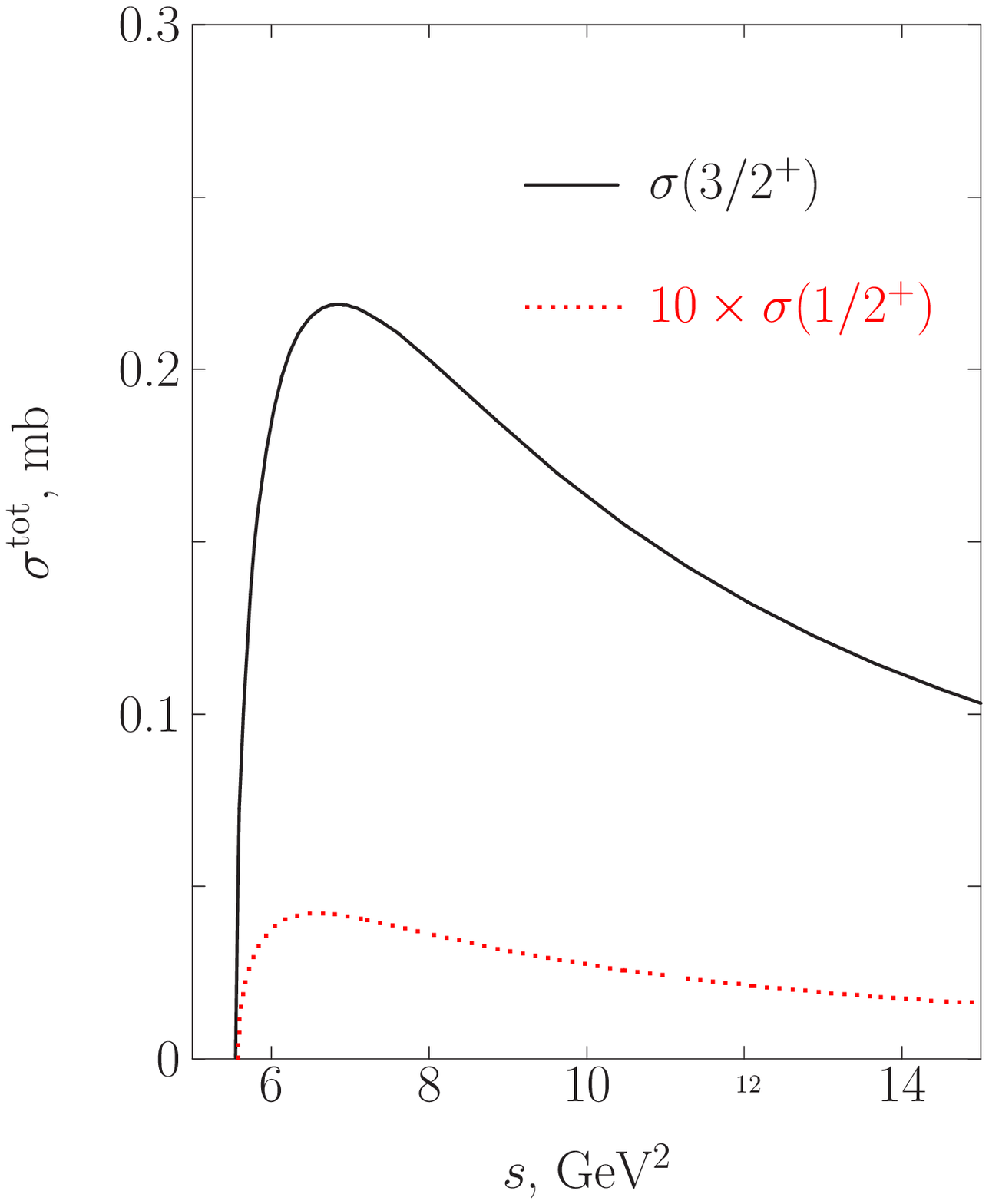}  
  \includegraphics[height=0.4\textheight]{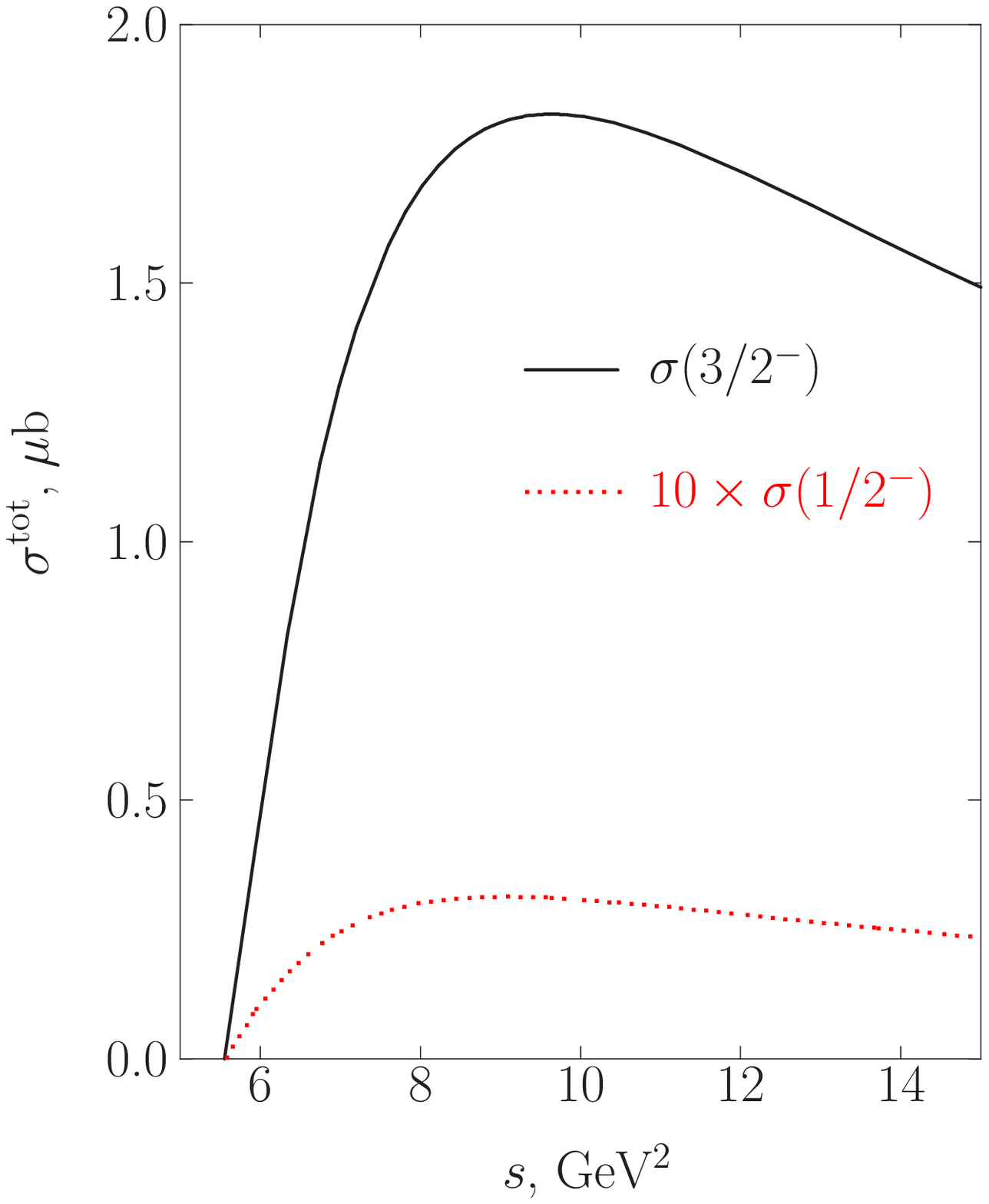}
  \caption{The total cross section for $\XXi^-$ production in $K^-n \to K^+\XXi^{--}$ reaction. The left panel is for positive and the right panel is for negative parity of the $\XXi^-$~baryon. The solid curves are for the $\XXi$ with the spin $J=\frac32$. The dotted curves are the results of Ref.~\cite{Hosaka} for the $\XXi$ with the spin $J=\frac12$.}
 \label{fig:C-section}
\end{figure} 

The estimated total cross section of the $\XXi$ production with spin $\frac32$ are displayed on Figure~\ref{fig:C-section}. We also compare our results with the results of Ref.~\cite{Hosaka} for the $\XXi$ with spin $\frac12$. One concludes that
\begin{itemize}
\item The total cross section for the same parity but different spin of the $\XXi$ is approximately 50--100 times larger for the spin $\frac32$ than for the spin $\frac12$.
\item Similarly to the case of spin $\frac12$ \cite{Hosaka} the cross section $\sigma(J^\pi=\frac32^+)$  is approximately two orders larger than  $\sigma(J^\pi=\frac32^-)$.
\end{itemize} 

It must be also stressed that according to (\ref{width}) the cross section is proportional to the $\Theta^+$ width, $\Gamma_{\Theta^+}$, which is unknown from experiment. The coupling constant $g_{K\Sigma\XXi}$ in Table~\ref{tab:1} was estimated from the ``average'' upper limit of the width, $\Gamma_{\Theta^+}<15$~MeV.  Further restrictions on the width come from the $K^+d$ total cross section, $\Gamma_{\Theta^+}<6$~MeV, \cite{Nussinov}, and from PWA of $K^+N$ scattering in the $I=0$ channel, $\Gamma_{\Theta^+}<1$~MeV, \cite{Arndt,Haidenbauer}. This means, that real value of the total cross section may be one order less than that given by Figure~\ref{fig:C-section}. 

\section{Conclusions\label{Conclusions}}
We estimate the upper limit of the total cross section for the $\XXi$ production in $\bar KN$ scattering employing $\XXi$ spin-parity $J^\pi=\frac32^+$ and $J^\pi=\frac32^-$. The estimate was done using the $s$ and $u$ pole diagrams with $\Sigma$ hyperon in the intermediate state, Figure~\ref{fig:diagrams}. We compare our results with the results of \cite{Hosaka} for the $\XXi$ with $J^\pi=\frac12^+$ and $J^\pi=\frac12^-$. It is shown that from the two cross sections with the same parity and different spins, the cross section for the spin $\frac32$ of the $\XXi$ is 50--100 times larger than that for the $\XXi$ with spin $\frac12$.   

\end{document}